\documentclass[journal,compsoc]{IEEEtran}
\usepackage{epsfig}
\usepackage{amssymb}
\usepackage{amsmath}
\usepackage{amsfonts}
\usepackage{amsthm}
\usepackage{mathtools}
\usepackage{alltt}
\usepackage{epstopdf}
\usepackage{algorithmic}
\usepackage[ruled]{algorithm}
\usepackage{multirow}
\usepackage{xcolor}
\usepackage{diagbox}
\usepackage{colortbl}
\usepackage{cite}
\graphicspath{ {image/} }
\usepackage{tabu}
\usepackage{subcaption}
\usepackage[justification=centering]{caption}
\usepackage{silence}
\begin{document}
	
	
	\title{Malware-on-the-Brain: Illuminating Malware Byte Codes with Images for Malware Classification}
	
	\author{Fangtian~Zhong,
		    Zekai~Chen,
		   Minghui~Xu,
		   Guoming Zhang,
		   Dongxiao Yu,
		Xiuzhen~Cheng,~\IEEEmembership{Fellow,~IEEE}
	\IEEEcompsocitemizethanks{
			\IEEEcompsocthanksitem Fangtian Zhong and Zekai Chen are with the Department of Computer Science, The George Washington University, Washington DC, USA. Emails: \{squareky{\_}zhong,zech{\_}chan\}@gwu.edu\protect\\

	Minghui Xu, Guoming Zhang, Dongxiao Yu and Xiuzhen Cheng (Corresponding Author) are with the School of Computer Science and Technology, Shandong University, Qingdao, 266237, China. Emails: \{mhxu,guomingzhang,dxyu,xzcheng\}@sdu.edu.cn\protect\\
		}
	}
	\markboth{This paper has been accepted by IEEE Transactions on Computers for final publication in an upcoming issue}%
	{\MakeLowercase{\textit{et al.}}}
	
	\IEEEtitleabstractindextext{
		\begin{abstract}
			  Malware is a piece of software that was written with the intent of doing harm to data, devices, or people. Since a number of new malware variants can be generated by reusing codes, malware attacks can be easily launched and thus become common in recent years, incurring huge losses in businesses, governments, financial institutes, health providers, etc. To defeat these attacks, malware classification is employed, which plays an essential role in anti-virus products. However, existing works that employ either static analysis or dynamic analysis have major weaknesses in complicated reverse engineering and time-consuming tasks. In this paper, we propose a visualized malware classification framework called VisMal, which provides highly efficient categorization with acceptable accuracy. VisMal converts malware samples into images and then applies a contrast-limited adaptive histogram equalization algorithm to enhance the similarity between malware image regions in the same family.  We provided a proof-of-concept implementation and carried out an extensive evaluation to verify the performance of our framework. The evaluation results indicate that VisMal can classify a malware sample within 4.0 ms and have an average accuracy of 96.0\%. Moreover, VisMal provides security engineers with a simple visualization approach to further validate its performance. 
		\end{abstract}
		
		
	}
	
	\maketitle
	
    \section{Introduction}
	\label{sec:introduction}
	
	\IEEEPARstart{M}{alware} classification refers to the process of grouping malware samples with similar features to effectively classify unknown malware. Features can be categorized as either static or dynamic, with the former being extracted based on a byte-code sequence, binary assembly instructions, or an imported Dynamic Link Library (DLL), and the latter on runtime file system activities, terminal commands, network communications, and function or system call sequences. Since malware samples in the same family normally have similar static or dynamic features, which could be extracted via machine learning algorithms, security vendors and researchers usually utilize them as useful clues for effective malware classification \cite{PEKTAS,multiclass}. 
		
	
However, malware feature extraction, whether static or dynamic, is a non-trivial task. The extraction of static features requires security professionals to have a good command of the structure of malware binaries. For instance, to extract an imported Dynamic Link Library in a Windows program, developers have to know the PE before parsing two structures in the order of DataDirectories followed by Import or Import Address Table \cite{malfox}. Additionally, it is even worse when encryption or compression is applied to malware binaries as reverse analysis is needed \cite{limsta}. On the other hand, although the extraction of dynamic features obviates the conundrums and limitations of static feature extraction, dynamic analysis can only collect the behaviors of specific execution paths since it is hardly possible to traverse all of them. Meanwhile, dynamic analysis is  time-consuming. For instance, some applications may have multiple event handlers that should be examined for dynamic feature extraction. Nevertheless, it can be insufficient to call just the one that contains a particular API invocation; instead, multiple event handlers may need to be triggered as a ``chain'' in a particular order with specific inputs \cite{tardyna}.

One can see from the above analysis that malware classification involves either very high knowledge barriers for security professionals or huge computing burdens for computers. To overcome the knowledge obstacles and improve malware classification efficiency, we have to address the following three critical challenges when designing malware classification schemes. First, since the ultimate purpose of malware classification is to precisely classify different malware samples, the extracted features should uniquely represent the samples and maintain the similarity for malware in the same family. Second, as the schemes are expected to be efficient, the time for extracting features and classifying malware samples should be limited to some degree that users can withstand. Third, even though security engineers may not have enough background knowledge about malware samples, the schemes should be easy to use and comprehend. Based on these considerations, we present a novel but simple visualized classification framework named VisMal in this paper, which aims at understanding the inner workings of malware families and disclosing the encoding art in the byte codes of malware binaries.  VisMal consists of three major components: Converter, Feature Engineer, and Classifier.  Converter takes charge of converting a malware sample into a 2-Dimensional grayscale image; Feature Engineer enhances the recognition of malware via strengthening the local contrast in the malware image regions and resizes the image to a smaller one to expedite the classification process; and Classifier is used to partition malware samples into their corresponding families.  
	 
	VisMal endeavors to greatly reduce the burdens on security engineers and the computing expenses on computers. Since existing works adopt either static or dynamic analysis approaches, which inevitably have drawbacks as mentioned above, it is necessary to turn our mind to pursue the essence of the malware families themselves. Note that many variants of malware samples are generated by reusing core codes \cite{codereuse,codereuse1,codereuse2,codereuse3}, one can use the similarity of instruction sequences to classify different types of malware samples. We provide an implementation of VisMal and conduct extensive experimental studies to evaluate its effectiveness in terms of accuracy, efficiency, and visualization. This study makes every effort to understand malware families and discover the underlying encoding art in byte codes. Our multi-fold contributions can be summarized as follows.
	
	\begin{enumerate}
		\item We propose VisMal, a visualized malware classification framework, to efficiently distinguish different types of malware samples while in the mean time maintaining high accuracy.   
		\item By improving local contrasts of the regions in converted images, we enlarge the discernity of malware samples, which elevates the accuracy from grayscale images without extra operations.
		\item When malware samples are compressed or encrypted, VisMal still demonstrates good performance on distinguishing different malware families. 
		\item VisMal provides a visual technique for security engineers without much knowledge about malware samples to readily validate the classification performance.
	\end{enumerate}

	The rest of the paper is organized as follows. Section~\ref{sec:background} introduces the most related work. Section~\ref{sec:framwork} details the  workflow and implementation of VisMal. Section~\ref{sec:evaluation} evaluates the performance of VisMal, and Section~\ref{sec:conclusion} concludes the paper with a discussion on future research.

	\section{Related Work}
	\label{sec:background}
	In this section, we provide an overview on popular existing  malware classification methods, considering whether they are  static analysis based or dynamic analysis based.
	
	\subsection{Malware Classification Based on Static Analysis}
	Since static features can be directly obtained from malware files without execution but they still possess certain discernibility, many approaches based on these features were proposed to classify different malware families, which are summarized as follows. 
	
	\textbf{Machine Learning-Based Approaches.} Jiang \emph{et al.} developed a $K$ Nearest Neighbour based (KNN) model for malware classification by taking an input vector carrying codes' semantic information that comes from sensitive opcode sequences \cite{opcodeknn}. Each sensitive opcode sequence is constructed by opcodes, application programming interfaces, STR instructions, and actions. All sensitive opcode sequences are sent to Doc2vec for the generation of sensitive semantic information, and the information is then used for classification in a pretrained KNN model. Fasano \emph{et al.} proposed a cascade learning method to distinguish different malware families by combining a KStar classifier for the discrimination of benignware and malware with a RandomForest classifier for the identification of different malware families \cite{quality}. The cascade learner makes use of totally 12 static values as features that are calculated from each malware, including cyclomatic complexity, weighted methods per class, bad smell method calls, wakelocks with no timeout, number of location listeners, number of GPS users, etc. Blanc \emph{et al.} trained a random forest classifier by leveraging the architectural and external attributes of malicious apps to identify different android malware families \cite{stametric}. To construct these attributes for a piece of malware, an app is initially decompiled with Apktool and converted into human-readable Smali codes. These codes are then parsed by a code quality tool and finally used for the extraction of 10 attributes (e.g., XML Parsers and Network Timeouts).
	
	\textbf{DNN-Based Approaches.} Jung \emph{et al.} proposed a convolutional neural network model based on byte information to classify malware samples \cite{byteseq}. The extraction of byte information  includes 3 steps: a malware binary file is first input to a disassembler, which returns the disassembled malware file; then byte length sequences are extracted from the disassembled malware file and used as parameters to a specific hash function; finally all hash values generated from the second step are used to construct a hash map, which is referred to as the byte information. Saxe \emph{et al.} presented a deep feed forward neural network trained on various features extracted by static analysis for binary malware classification  \cite{binarydp}. The features include system library import functions, ASCII printable strings, PE metadata fields in the executable, as well as byte sequences from the raw codes. They are concatenated to form a 1024-dimensional feature vector, which is fed into a four-layer feed forward network for training and further classification. Davis \emph{et al.} developed an approach that employs a convolutional neural network to identify malware \cite{disadp}. The raw disassembled codes of malware are processed to generate a vector of fixed-length features. This approach extracts the individual x86 processor instructions with diverse lengths and then apply padding or truncation to create fixed length features. The disassembled codes are also parsed to extract system library import functions to further extend the feature vector size.  
	
	\textbf{Other Approaches.}  Kinable \emph{et al.} investigated malware classification based on call graph clustering \cite{callclus}. Each malware can be represented as multiple call graphs that consist of local functions and system library functions as vertexes and caller-callee relationships as edges. Malware programs in the same family should have  minimized number of mismatchings in their graphs. To compute malware mismatchings, they adopted the graph edit distance that is the minimum number of elementary operations required to transform one graph to another. Mirzaei \emph{et al.} proposed a characterization approach for malware families based on common ensembles of sensitive API calls \cite{AndrEnsemble}. The overall architecture of the proposed system goes through five steps: it first computes fuzzy hash values of all functions extracted from the apps in the same family; then it creates call graphs and uses the fuzzy hash values to replace the nodes (functions); next,  it merges the hashed call graphs and all edges to form an aggregated graph and assigns a unique weight to each edge; following that it inspects the results from the frequency of similar methods and the frequency of the methods' invocations in the similar methods within a family, which are used to create feature vectors at last for classification. Zhang \emph{et al.} implemented a prototype system named DroidSIFT, a novel semantic-based approach to classify Android malware via dependency graphs \cite{semantics}. They generated dependency graphs by discovering entry points and analyzing call graphs. Following the generation of dependency graphs, they leveraged both forward and backward dataflow analysis to explore API dependencies and uncover constant parameters. DroidSIFT hence uses dependency graphs to build graph databases and classifies a malware sample by producing graph-based feature vectors and then performing graph similarity queries.

	   \begin{figure*}[!htb]
		\centering
		\includegraphics[scale=0.6]{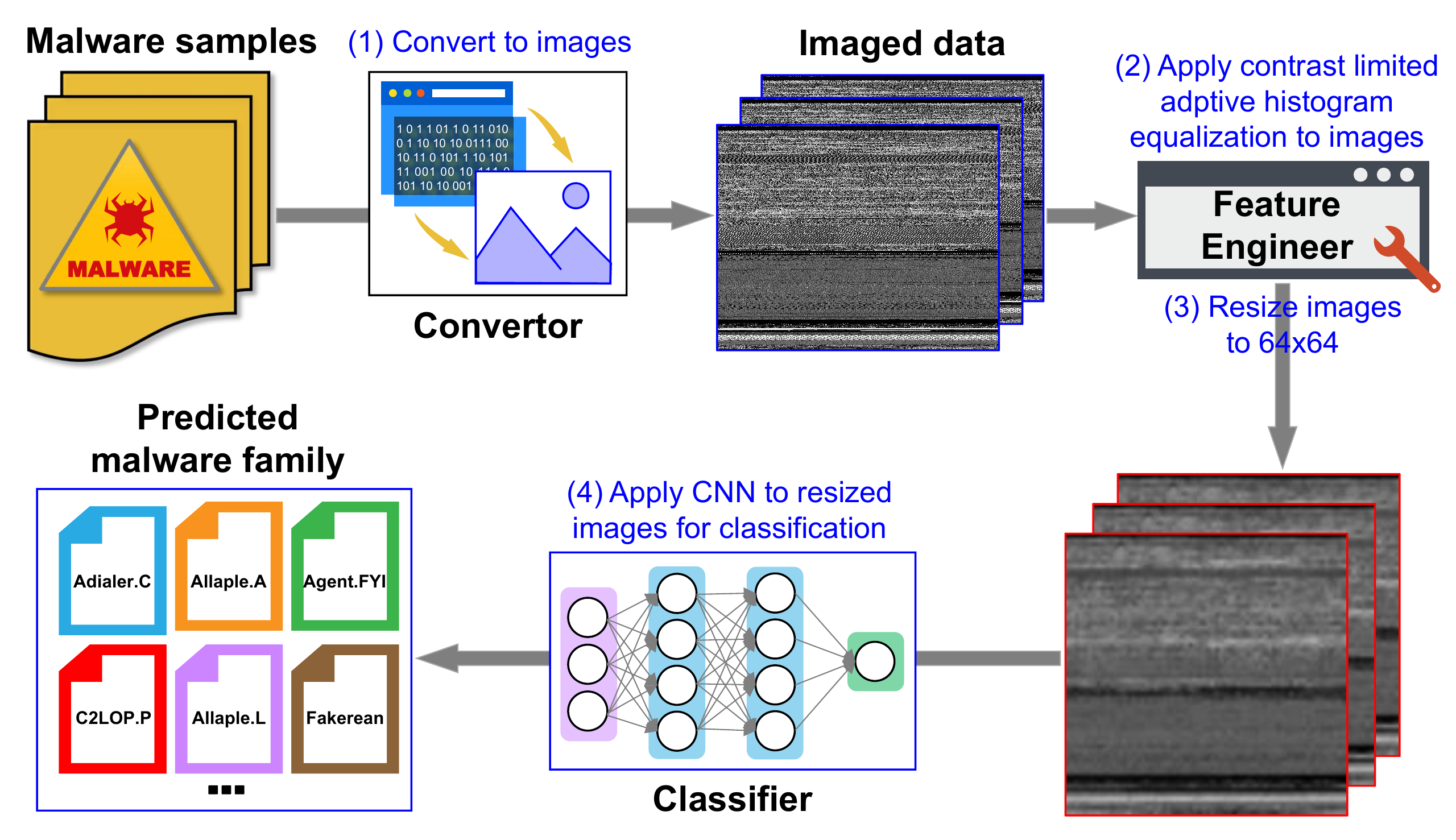}
		\caption{Overview of the VisMal Framework}
		\label{fig:vismal}
	   \end{figure*}
	
	\subsection{Malware Classification Based on Dynamic Analysis}
	 In contrast to static analysis, which derives properties from a program's text, dynamic analysis obtains properties by examining a running program. Approaches based on dynamic analysis are important complements to improve detective capability when static features are hard to extract due to obfuscations. In this subsection, we summarize major malware classification works based on dynamic analysis.
	
	\textbf{Machine Learning Based Approaches.} Sethi \emph{et al.} developed a machine learning based malware analysis framework for classification \cite{novelframe}. All malware samples are executed and analyzed on top of the Cuckoo Sandbox, which examines their run-time behaviors and returns behavioral reports that memorize their activities. The reports are further processed by a feature extraction and selection module before sent to machine learning models such as Decision Tree and Random Forest for final classification.  In \cite{candyman}, Martin \emph{et al.} implemented a tool termed CANDYMAN to classify Android malware by combining dynamic analysis and Markov chains. In their experiment,  DroidBox was employed as the dynamic analysis tool to generate a state sequence representation for each malware. The state sequence representation is then used to model the transition probabilities between consecutive states. Once all transition probabilities are computed, they are input to a collection of machine learning algorithms (i.e., SMOTE, Random Over Sampler, Random Under Sampler, etc.) for further classification. Anderson \emph{et al.} designed a malware classification algorithm based on graphs constructed from dynamically collected instruction traces of target execuables under the Ether system  \cite{graphmal}. Each instruction is denoted as a vertex in a graph, an edge between two nodes represents the transition from one instruction to another,  and the transition probabilities are estimated by all collected traces. One can acquire a similarity matrix with each entry describing the similarity between an executable and a malware class, which is obtained by employing graph kernels, and then send the matrix to a support vector machine for classification.

	\textbf{DNN-Based Approaches.} Dahl \emph{et al.} trained a neural network with three hidden layers by feeding it with system API calls as well as null terminated strings extracted from the process memory \cite{neuralproj}. A random projection technique was used to reduce the features' dimensionality to the degree that is manageable by the neural network. Tobiyama \emph{et al.} developed a recurrent neural network (RNN) to extract features from a detailed representation of malware process behaviors and used a convolutional neural network (CNN) to classify the outputs from the RNN \cite{processbe}. 
	Each API call represents an operation of the process, multiple API calls represent an activity, and multiple activities represent a behavior of the process. In each operation, critical information such as process name, ID, event name, and path of current directory is recorded. Zhang \emph{et al.} proposed a low-cost feature extraction approach and an effective deep neural network architecture for accurate and fast malware classification \cite{featureengine}. They executed PE files inside virtual machines and used API hooks to monitor the API call traces. They also utilized a feature hashing trick to encode the API call arguments associated with the API name for each API call. The deep neural network architecture applies multiple Gated-CNNs to transform the encoding into particular feature representations that are further processed by bidirectional LSTM (long-short term memory networks) to learn the sequential correlations among API calls. 
	
    \textbf{Other Approaches.} Vij \emph{et al.} proposed a novel graph signature based malware  classification mechanism \cite{gramac}. The graph signature uses sensitive application programming interfaces to capture the control flow that builds a bridge between sensitive APIs and the respective incidents. A dataset of graph signatures for well-known malware families is then created. A new application's graph signature is compared with the signatures in the dataset, and the application is classified into the corresponding malware family. Park \emph{et al.} developed a malware classification algorithm called FACT via finding the maximum common subgraphs in directed function call graphs \cite{automat}. Each graph is constructed by capturing the uses of the system functions along with the parameter values when the malware program is executed in a sandboxed environment. Since  malware programs belonging to the same family generally share similar behaviors by calling certain system functions, i.e., file system modification, new process spawning, network connectivity checking, registry modification, 
    etc., their subgraphs tend to be similar to some extent. Hsiao \emph{et al.} proposed an analysis scheme to group Android malware based on their dynamic behaviors, and to identify the behaviors of a malware family. They also applied phylogenetic tree, significant principal components, and dot matrix on different malware families to demonstrate their behavioral correlations. The proposed methods can automatically discover similar behaviors of different malware groups, extract the characteristics of each malware group, and provide security experts with the visualized information based on runtime behaviors \cite{behagroup}.
    
\subsection{Summary}
   
    In this paper, we present VisMal, which converts malware binaries into 2-Dimensional grayscale images and utilizes a contrast-limited adaptive histogram equalization algorithm to expand the difference in all image regions such that it is easier for our classifier to capture the similarity of malware samples in the same family. Previous  works such as \cite{multiclass,quality,tardyna,PEKTAS}  perform less effective when reverse analysis is an obstacle and time is a constrained factor. VisMal cleverly bypasses the obstacles of reverse analysis by making effort to understand the properties of malware families, and discover the underlying similarity in byte codes. Additionally, VisMal processes a malware program as an image, significantly reducing the waiting time for analysis. These two features make VisMal more practical and efficient. VisMal is also different from the works such as \cite{texture,malimg,multi,YUAN} that process malware as images, which focus on statistical features of the converted image -- VisMal pays attention to the understanding of the correspondence between byte codes and pixels. Besides, VisMal does not need to intentionally pick out particular statistical features for classification.
    
From the previous analysis, one can see that malware classification based on static features  distinguishes different malware samples mainly by grouping their text and statistical information extracted from malware files. Text information includes opcode sequences, system library import functions, and ASCII printable strings, while statistical information calculated by various tools involves cyclomatic complexity, number of GPS users, and network timeouts, etc. Different from static analysis based methods, VisMal does not require the disasembling of the malware samples and the help of any tool. On the other hand, malware classification based on dynamic  features deals with the running programs on top of virtual machines and examines malware samples by their programs' activities, instruction traces, API call traces, control flows, etc. In contrast to such methods, VisMal does not need to execute programs and check each execution trace to assure the reliability of classification. Besides, the selection of a distinct collection of features to uniquely represent a malware sample  is not required by VisMal.
    
    \section{The Proposed VisMal Framework}	
    \label{sec:framwork}

    \subsection{Overview of VisMal}
	\label{sec:design}
	
	The three-fold objectives of VisMal include: i) efficiently classify malware samples into their respective families, ii) reduce security engineers' burdens by providing automatic classification tools, and iii) offer an image-based validation method for identifying the effectiveness of classification. The framework of VisMal is presented in Fig.~\ref{fig:vismal}. One can see that VisMal consists of 3 components: \emph{Converter}, \emph{Feature Engineer}, and \emph{Classifier}. Converter converts a malware binary into an image. Feature Engineer processes the image by a contrast-limited adaptive histogram equalization algorithm to increase the local contrast in different image regions and then resize the processed image to a smaller fixed-size one that can accelerate the process of classification. 
	Classifier employs a shallow convolutional neural network-based model.
	 It enables parameter sharing, a technique that applies the same small filter to all elements of the input and captures the correlation of different elements,  to allow CNN to efficiently extract the local features in an image. 
	
The workflow of VisMal is described as follows.  All malware samples, whether in the training dataset or in the test datasets, need to be converted to a grayscale image according to the following procedure: a malware example is first sent to  Converter for numeric conversion, which outputs an array consisting of numbers with scope between 0 and 255; then the array is reshaped according to its file size (see Table \ref{tbl:tab1}); following the reshaping is the feature editing and resizing processes by Feature Engineer. The output of Feature Engineer is an image array scaled to 64x64. Then it is concatenated with the corresponding malware family label. The data is randomly divided into a training dataset and test dataset by StratifiedKFold. This is a variation of KFold that returns stratified folds made by preserving the percentage of samples for each class \cite{kfold}. The training dataset is used to train Classifier. Once the classifier is well trained, one can use it to characterize the malware samples in the test dataset.

	\begin{table}[!htb]
		\captionsetup[table]{skip=10pt}
		\caption{Correspondence between the  file size of a malware sample and the converted image width}
		\label{tbl:tab1}
		\centering
		\begin{tabular}{|m{3.5cm}<{\centering}|m{2cm}<{\centering}|m{2cm}<{\centering}|}
			\hline
			File Size  &   Width & Height \\ 
			\hline
			$\le$ 10KB      &   32  &  (0, 312]  \\ 
			\hline
			10KB-30KB  &   64  &   (156, 468]     \\ 
			\hline
			30KB-60KB  &   128 & (234, 468]  \\ 
			\hline
			60KB-100KB &  256  &   (234, 390]  \\ 
			\hline
			100KB-200KB &  384 & (260, 520]\\
			\hline
			200KB-500KB & 512 & (390, 976] \\
			\hline
			500KB-1000KB& 768 & (651, 1302] \\
			\hline
			1000KB $\leq$&1024 &(976,$\infty$) \\
			\hline
		\end{tabular}	
	\end{table}

	\subsection{The Components of the VisMal Framwork}
	\label{sec:implementation}
	
	In this section, we detail the implementations of the three components of VisMal, namely \emph{Converter}, \emph{Feature Engineer}, and \emph{classifier}.
	\begin{figure}[!htb]
		\centering
		\includegraphics[scale=0.4]{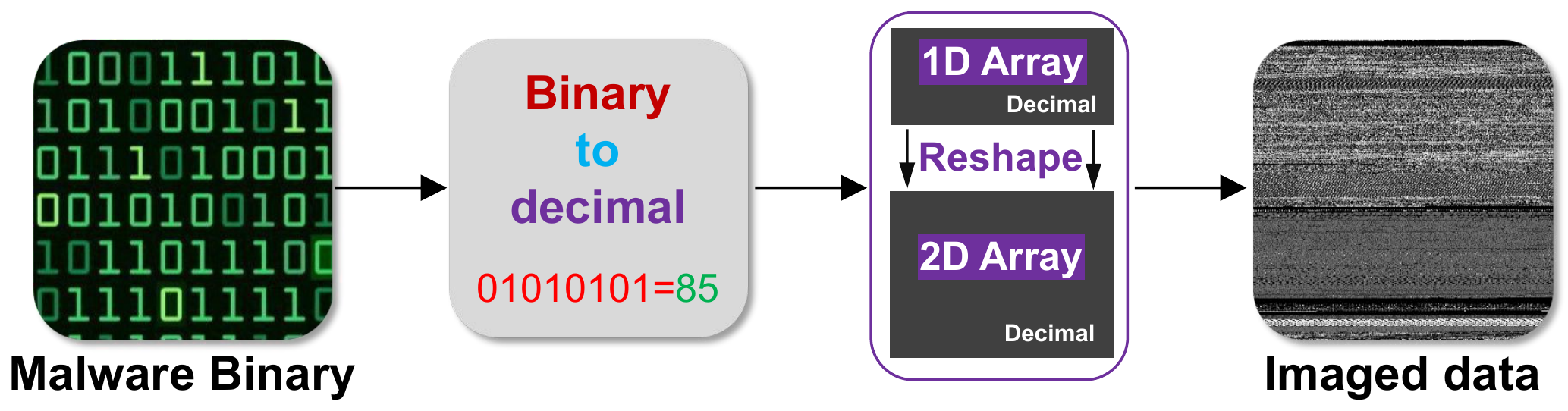}
		\caption{Converter}
		\label{fig:converter}
	\end{figure}
	\subsubsection{Converter}\label{sec:extractor}
    Fig. \ref{fig:converter} demonstrates the procedure of numeric conversion by Converter for a malware example. To convert a malware binary into a grayscale image, Converter sequentially reads the binary data in bytes, converts each byte into a decimal number ranging in [0-255] \cite{texture}, then saves the number in a one-dimensional array. For instance, '0110000' is converted to 96. Every number in the array corresponds to a pixel in a grayscale image. As the resolution of a grayscale image has width and height dimensions but different malware samples vary in their file sizes, one should consider such a situation to accept diverse options of width and height. In this paper, we take a simple approach to reshape the image data (array) by following the recommended fixed-width with a variable height according to its file size, which is correspondingly transformed into a 2-Dimensional grayscale image conforming to portable network graphics (.png extension) via a built-in function, i.e., imwrite or imsave, in cv2 library or PIL library. The recommended fixed image widths for distinct malware file sizes are given in Table \ref{tbl:tab1} \cite{malimg}.

	\subsubsection{Feature Engineer}\label{sec:engineer}
	As suggested in  \cite{codereuse,codereuse1,codereuse2,codereuse3}, many variants of malware samples are generated by reusing core codes. Therefore one can use the similarity of instruction sequences to classify different malware samples. Since an instruction typically has multiple bytes and each byte corresponds to one pixel, the instruction is converted into several pixels by Converter. Thus locating similar instruction sequences from different malware samples is equivalent to identifying regions with similar pixel values in their corresponding images. Nevertheless, similar instruction sequences from different malware samples belonging to the same malware family may exist at different positions of their files. To overcome this challenge, we resort to the image contrast to identify similar pixel regions -- image regions produced by similar instruction sequences should have similar contrast values.  To further improve the accuracy of Classifier, the capture of similar pixel values in different malware images should be promoted. Since the usable data of an image is normally represented by close contrast pixel values  \cite{he}, increasing the contrast of an image benefits classification. Therefore, Feature Engineer adjusts pixel intensities in each image region to enhance the contrast for a malware image by applying a contrast-limited adaptive histogram equalization algorithm. This algorithm maps the narrow range of input pixel values for an image region to the whole range of the pixel values for an image, i.e., 0-255. Through this adjustment, the intensities can be better distributed on the histogram\footnote{a histogram correlates to the frequency distribution of pixel values for an image region.} for each image region by effectively spreading out the most frequent intensity values, which allows for areas of lower contrast to gain higher contrast and improves the contrast for the entire image. High contrast enhances the possibility of Classifier to identify similar pixel values in images.

\begin{figure}[!htb]
		\centering
		\includegraphics[scale=0.35]{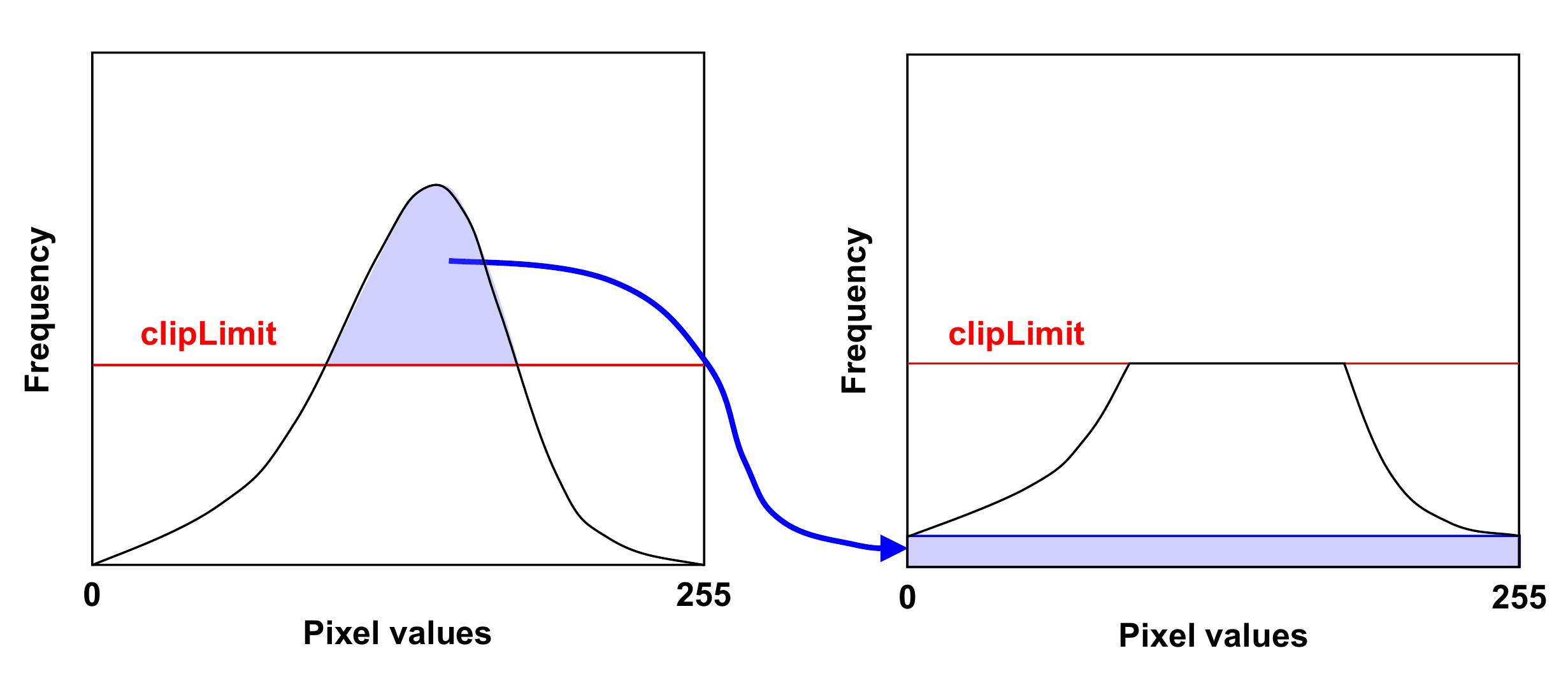}
		\caption{Clipping}
		\label{fig:clipping}
	\end{figure}

	Figuratively speaking, contrast-limited adaptive histogram equalization seems to be equivalent to signal transform  that happens in biological neural networks so as to maximize the output firing rate of a neuron which is a function of the input statistics. The equalization algorithm contains 4 phases: Division, Cumulation, Clipping, and Transformation. Division divides an image into $a \times b$ small regions where $a$ and $b$ are respectively the numbers of pieces split up for the width and height of the image. A cumulation is a cumulative frequency distribution function for an image region, whose computation is shown in \eqref{cumulation}:
	\begin{equation} \label{cumulation}
	cdf(i)  =  \sum\limits_{j = 0}^i {{n_j}} ,0 \le i < L 
	\end{equation}
	where $L$ is the total number of gray levels (typically 256), and $n_j$ is the number of times  the pixel value $j$ appears in an image region. Note that $n_j$ is also called the frequency of $j$. As shown in Fig.~\ref{fig:clipping}, if the frequency of a pixel value is above the threshold $clipLimit$, Clipping is applied to assign a random value in $[0,255]$ to some of such pixels such that no pixel can have a frequency higher than $clipLimit$ (see Fig. \ref{fig:clipping}); correspondingly, the histogram is adjusted to a new frequency distribution for the image region. 	
	Transformation transforms the input pixel values into output pixel values by adopting different transformation methods according to their positions in the malware image. 
		
	\begin{figure}[!htb]
		\centering
		\includegraphics[scale=0.4]{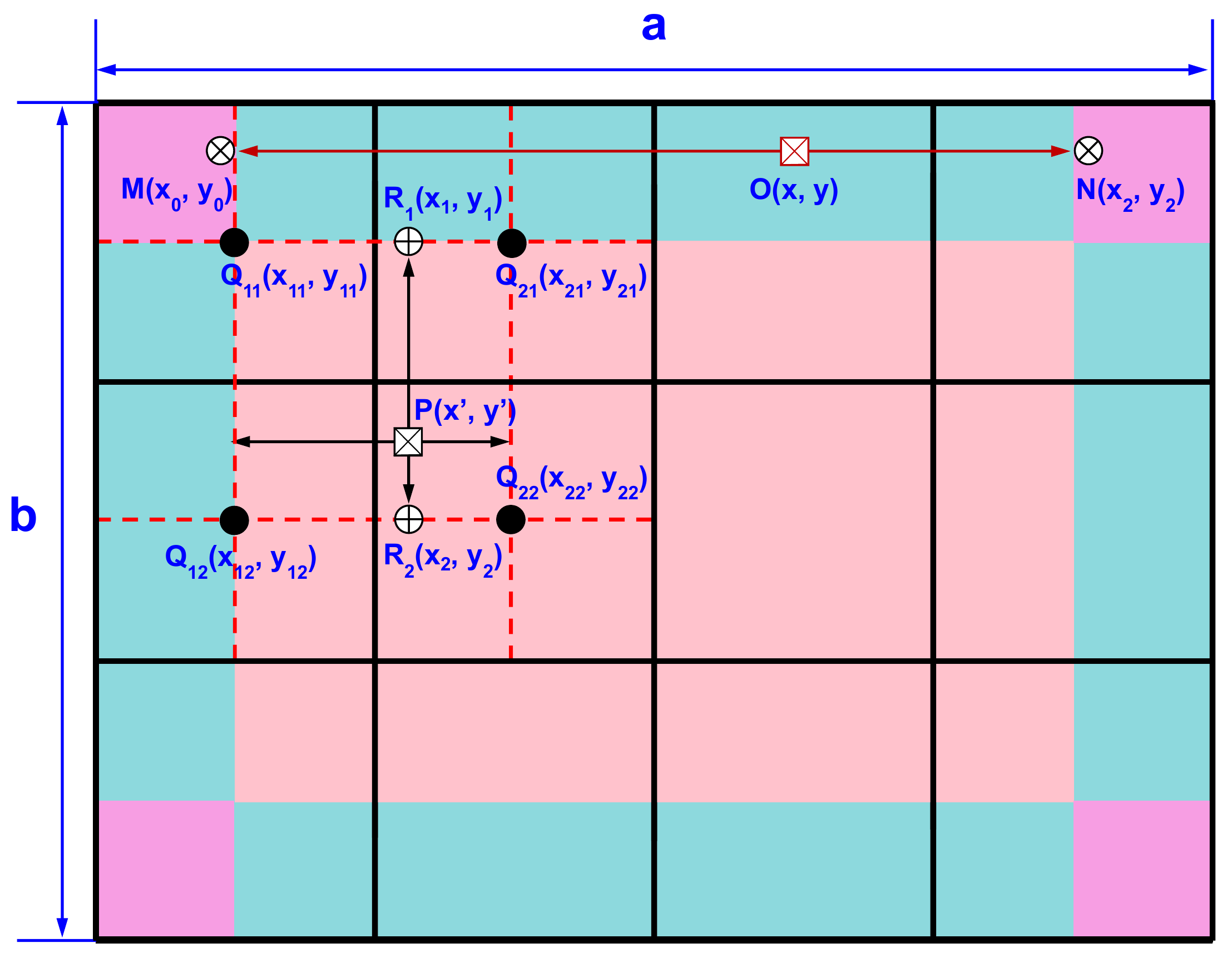}
		\caption{Transformation of Input Pixels}
		\label{fig:transformation}
	\end{figure}
	
	We use Fig.~\ref{fig:transformation} as an example to illustrate the Transformation phase. In Fig.~\ref{fig:transformation}, the image is partitioned into $4 \times 3$ equal-sized rectangular regions. The center of each region is represented by a black dot, and each boundary region is partitioned into 4 smaller rectangles by the center lines, resulting in areas with three different colors (see Fig.~\ref{fig:transformation}). Each area includes the upper and left borders as well as the outermost border of the image if it is a boundary one. The input pixel values in the rectangle areas shaded purple, sky blue, and pink are transformed by histogram equalization, linear interpolation, and bilinear interpolation, respectively. More specifically, according to \cite{ahe},  pixels in the purple areas and the center pixel of each region  are transformed by histogram equalization; each pixel in the sky blue areas is transformed based on two collinear pixels in the two neighboring purple areas by linear interpolation; and pixels in the pink area that lie between two collinear center pixels are transformed by linear interpolation while other pink pixels are transformed by bilinear interpolation based on the four neighboring center pixels. For example, consider the points $M, N, Q_{11}, Q_{12}, Q_{21}, Q_{22}, O, R_1, R_2, P$ in different areas of Fig.~\ref{fig:transformation}, where
	the value of the $x$-coordinate for each pixel point is the input pixel value, and that of the $y$-coordinate is the output pixel value. 
	The point $M$ in the purple area and the center points $Q_{11}, Q_{12}, Q_{21}$, and $Q_{22}$ of the four regions in the upper-left of the image are transformed by histogram equalization according to the following transformation function
	\begin{equation}\label{transfunc}
	y = h_R(x)  =  round(\frac{{cdf({{x}}) - cd{f_{\min }}}}{{cd{f_{\max }} - cd{f_{\min }}}}{\rm{ \times }}(L - 1)) 
	\end{equation}
	 over all the input pixel values in the image region  the point lies in to produce its output pixel value, where ${cd{f_{\min }}}$ is the minimum non-zero value of the cumulative distribution function calculated in the Cumulation phase while ${cd{f_{\max }}}$ gives the maximum value.
	
	Note that \eqref{transfunc} maps a narrow range of the input pixel values to the full range of the image pixel values. However, this equation may overamplify the small amounts of noises in largely homogeneous image regions \cite{clahe}. For example, if most pixel values are big numbers, and noises' pixel values are small numbers, the transformation of the noises may fall into the same pixel value such that the noises are overamplified as the frequency of the corresponding pixel values increases. Clipping is thus adopted to equally redistribute the strongly peaked pixel values in the histogram in order to limit the amplification of the noises when the frequency of the pixel values surpasses a prescribed threshold $clipLimit$. 

On the other hand, a point $O$ in the sky blue area located at the horizontal line $MN$, where $M$ and $N$ with pixel value $x_0$ and $x_2$, respectively, are points in the left-upper and right-upper corner purple areas of the image, can be linearly interpolated as follows:  
%
	\begin{equation}\label{eq:linear}
 y = \frac{{{x_2} - x}}{{{x_2} - {x_0}}}h_{R_M}({x_0}) + \frac{{x - {x_0}}}{{{x_2} - {x_0}}}h_{R_N}({x_2})
 \end{equation}
	
	Besides, point $P$, an arbitrary point in the pink area, is bilinearly interpolated: its input pixel value $x'$ is transformed to the output pixel value $y'$ based on the input pixel values and output pixel values of the four neighboring center points ($Q_{11}, Q_{12}, Q_{21}$, and $Q_{22}$) according to equations  \eqref{eq:R1},  \eqref{eq:R2}, and \eqref{eq:P}. Specifically, \eqref{eq:R1} and  \eqref{eq:R2} respectively compute the output pixel values of $R_1$ and $R_2$, which are then employed by \eqref{eq:P} to generate the output pixel value of $P$.
	\begin{align}
 y_1 & = \frac{{{x_{21}} - x_1}}{{x_{21} - {x_{11}}}}h_{R_{Q_{11}}}(x_{11}) + \frac{{x_1 - {x_{11}}}}{{{x_{21}} - {x_{11}}}}h_{R_{Q_{21}}}(x_{21}) \label{eq:R1}
 \\
 y_2 &= \frac{{{x_{22}} - x_2}}{{{x_{22}} - {x_{12}}}}h_{R_{Q_{12}}}(x_{12}) + \frac{{x_2 - {x_{12}}}}{{{x_{22}} - {x_{12}}}}h_{R_{{Q_{22}}}}(x_{22}) \label{eq:R2}
    \\
 y' &= \frac{{{y_2} - h_{R_P}(x')}}{{{y_2} - {y_1}}}y_1 + \frac{{h_{R_P}(x') - {y_1}}}{{{y_2} - {y_1}}}y_2 \label{eq:P}
 \end{align}

	\begin{figure*}[!htb]
		\centering
		\includegraphics[scale=0.5]{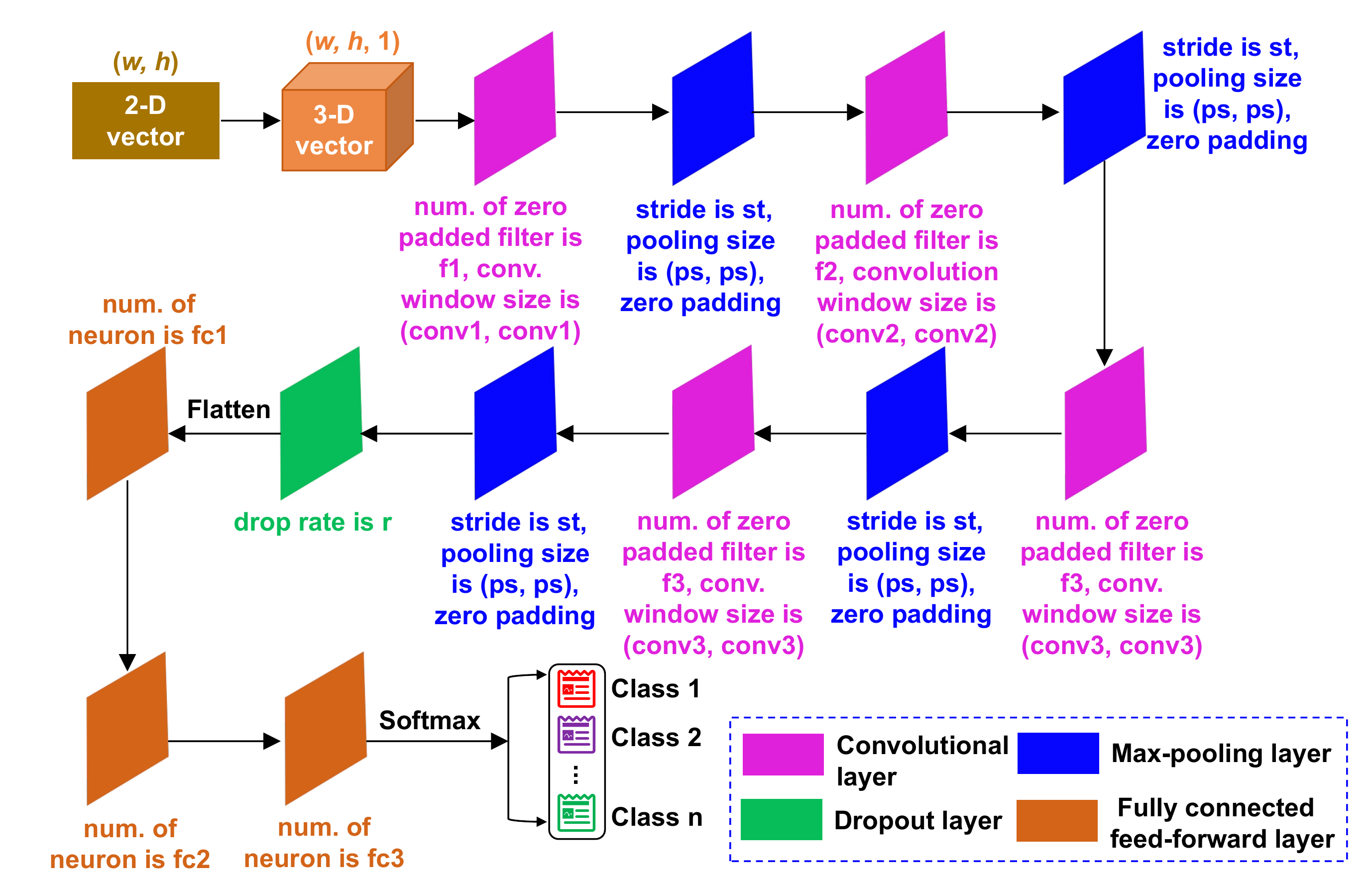}
		\caption{Network Structure of Classifier}
		\label{fig:classifier}
	\end{figure*}

	One can see that Transformation makes the pixel values inter-related,  enhancing the contrast of the image. Next, the transformed image is resized with a shape $(s, s)$ and then input to Classifier.

	\subsubsection{Classifier}
	Classifier estimates the probabilities at which a malware sample is classified to a certain family by employing a CNN algorithm. The \emph{classifier} employed by our study for the dataset presented in Section~\ref{sec:evaluation} is shown in Fig. \ref{fig:classifier}, which contains 12 layers: four convolutional layers, four max-pooling layers, one dropout layer, and three fully connected feed-forward layers, and are detailed in the following paragraph. CNN is adopted here because it has demonstrated promising classification results in many fields, particularly in image processing, which could precisely distinguish different malware families for our purpose. 

  The \emph{classifier} in Fig. \ref{fig:classifier}  is constructed with a training dataset that consists of a collection of different types of malware programs. The input is a 2-D vector with a shape of ($w$, $h$), where $w$ and $h$ are respectively the width and height of an image. This vector is first reshaped into a 3-D vector with a shape of $(w, h, 1)$ to meet the shape requirements of the first convolutional layer. The first convolutional layer contains $f1$ zero-padded filters and a convolution window with a shape of $(conv1, conv1)$.  One can get the output of this layer with more channels in-depth, \emph{i.e.}, $X_{1}^C$ has the shape of $(w, h, f1)$. To avoid over-fitting, all convolutional layers would have an L2 regularization penalty with a regularization factor $l2$ on the layer's kernel. Following the first convolutional layer is the first max-pooling layer with a stride $st$, a pooling size $(ps, ps)$, and zero paddings. All following max-pooling layers have the same settings. The output of the first max-pooling layer is input to the second convolutional layer that contains $f2$ zero-padded filters and a convolution window shaped $(conv2, conv2)$. The output from the second convolutional layer is sent to the second max-pooling layer, followed by the third convolutional layer possessing $f3$ zero-padded filters and a convolution window shaped ($conv3, conv3$), and the third max-pooling layer. The output of the third max-pooling layer is processed by the fourth convolutional layer and the fourth max-pooling layer, which are configured similarly to the previous convolutional layer and max-pooling layer. The dropout layer with a drop rate $r$ following the fourth max-pooling layer plays the role of promoting generalization. The output of the dropout layer is flattened and then sent to the first fully-connected feed-forward layer with $fc1$ neurons, followed by the second and third fully-connected feed-forward layer with $fc2$ and $fc3$ neurons, respectively. Note that  $fc3$ is the number of malware families our dataset has. Also, note that the third fully-connected feed-forward layer contains a softmax activation function that gives the probability of the malware being classified to a certain family, and the calculation is shown in \eqref{softmax}.  
	\begin{equation}\label{softmax}
	X_{{i_{j'}}}^C = \frac{\exp \left( {X_{{{\rm{i}}_j}}^C} \right)}{{\Sigma _k}\exp \left( {X_{{{\rm{i}}_k}}^C} \right)}.
	\end{equation}
where $X_{{i_{j}}}^C$ is the output value of the $i$th layer that is relevant to the $j$th malware family.
	The loss function of the \emph{classifier} is defined in \eqref{classifier}, where $y_i$ is the true label of a malware sample, and $\hat{y_i}$ 
	 is the predicted label for the malware sample.
	\begin{equation}\label{classifier}
 Loss =  - \sum\limits_{i = 1}^{f3} {y_i^{}\log \hat{y_i}  }
	\end{equation}
	
	To precisely distinguish different types of malware in the dataset, $Loss$ should be minimized about the weights and parameters of  the \emph{classifier}. Minimizing $Loss$ would improve the predicted probability of malware that belongs to its corresponding family. When employing  the \emph{classifier}  to test a malware sample, the input is a malware image and the output is the malware family the image (malware) belongs to.

	\section{Evaluation}
	\label{sec:evaluation}
	The metrics to measure the performance of VisMal include  \emph{accuracy}, \emph{efficiency}, and \emph{visualization}. Let $C_{i,j}$ be the number of malware samples in family $i$ that are classified to family $j$ by Classifier, where $i,j\in \{1,2,\cdots, N\}$ and $N$ is the total number of malware families. Then one can utilize  TP, FN, FP, and TN, the four common machine learning parameters to define the performance metrics of VisMal.  More specifically, for a malware family $i$, $T{P_i}$ provides the number of correctly predicted samples belonging to the family $i$, represented by $T{P_i}$=${C_{i,i}}$; $T{N_i}$ is the number of correctly predicted samples belonging to other families, represented by $T{N_i} = \sum\limits_{{\rm{j}} = 1,j \ne i}^N {{C_{j,j}}}$; $F{N_{\rm{i}}}$ is the number of misclassified samples that belong to family $i$, represented by $F{N_{\rm{i}}} = \sum\limits_{j = 1}^N {{C_{i,j}} - {C_{i,i}}}$; and $F{P_{\rm{i}}}$ is  the number of samples misclassified into family $i$, represented by $F{P_{\rm{i}}} = \sum\limits_{j = 1}^N {{C_{j,i}} - {C_{i,i}}}$. Then, the metrics to examine the performance of Classifier can be formally denoted as:
	\begin{enumerate}
		\item The \emph{accuracy} is the ratio of correctly predicted malware samples over all malware samples:
		\begin{equation}\label{eq:accuracy}
		accuracy = \frac{{TP}_i+{TN}_i}{{TP}_i+{FN}_i+{FP}_i+{TN}_i}\cdot
		\end{equation}
		\item The \emph{precision} for a malware family $i$ is the ratio of correctly classified samples over the total samples classified into this family:
		\begin{equation}\label{eq:precision}
		precision = \frac{{TP}_i}{{TP}_i+{FP}_i}\cdot
		\end{equation}
		\item The \emph{recall} for a malware family is the ratio of correctly classified samples over the total samples belonging to this family:
		\begin{equation}\label{eq:recall}
		recall = \frac{{TP}_i}{{TP}_i+{FN}_i}\cdot
		\end{equation}
		\item The \emph{F1} score for a malware family $i$ is the balanced average of precision and recall:
		\begin{equation}\label{eq:F1}
		F1\ score = \frac{2*(recall*precision)}{recall+precision}.
		\end{equation}
	\end{enumerate}
	
	We also examine the efficiency of our VisMal framework. As VisMal is a time-sensitive system, its efficiency, denoted by \emph{MPE}, can be measured by the CPU processing time per malware sample that counts only feature extraction and classification. Let \emph{cputime} be the total clock ticks used to process $num\_files$ malware samples. We have 
	\begin{equation}\label{eq:efficiency}
	MPE = \frac{cputime}{num\_files}.
	\end{equation}
	
	Besides accuracy and efficiency, we also employ visualization to validate the performance of VisMal, providing security engineers with a convenient way to further analyze malware samples.
	
	\subsection{Experiment Setup}
	
	\textbf{Equipment and Dataset.} The PC employed by VisMal is equipped with 12 processors, 6 kernels, and an installed RAM with a 32.0GB available memory running the 64-bit Windows 10 operating system. Each processor is configured with Intel(R) Core(TM) i7-8750 CPU @2.20GHz, 2201Mhz. The versions of Keras, Tensorflow, Pillow, and OpenCV to develop our framework are 2.2.3, 2.4.0, 6.2.1, and 4.5.1.48, respectively. Additionally, we adopt the Malimg dataset released by the Vision Researccomprisesh Lab at the University of California, Santa Barbara, as training and test data \cite{dataset}. This dataset comprises 9339 malware samples in total, which belongs to 25 malware families with a varying number of malware files per family, as detailed in Table \ref{tbl:tab2}.
	
	\begin{table}[!htb]
		\captionsetup[table]{skip=10pt}
		\caption{Malimg Dataset}\label{tbl:tab2}
		\centering    
		\begin{tabular}{|m{2cm}<{\centering}|m{3cm}<{\centering}|m{2cm}<{\centering}|}
			\hline
			number & Family  & num\_files \\ 
			\hline
			1.    & Adialer.C & 122 \\   
			\hline
			2.    & Agent!FYI  & 116 \\ 
			\hline
			3.    & Allaple.A  & 2949 \\  
			\hline
			4.     &   Allaple.L & 1591\\  
			\hline
			5.     &  Alueron.gen!J & 198 \\ 
			\hline
			6.   &  Autorun.K  & 106 \\ 
			\hline
			7.  & C2Lop.gen!g  & 200 \\ 
			\hline
			8.   & C2Lop.P  & 146 \\ 
			\hline
			9.   & Dialplatform.B  & 177 \\ 
			\hline
			10.   &  Dontovo.A  & 162 \\
			\hline
			11.   &  Fakerean  & 381 \\ 
			\hline
			12.   & Instantaccess  & 431 \\
			\hline
			13.   & Lolyda.AA 1  & 213 \\ 
			\hline
			14.   & Lolyda.AA 2  & 184 \\ 
			\hline
			15.   &  Lolyda.AA 3  & 123 \\ 
			\hline
			16.   & Lolyda.AT & 159 \\
			\hline
			17.  & Malex.gen!J  & 136 \\
			\hline
			18.   & Obfuscator.AD  & 142 \\ 
			\hline
			19.   &  Rbot!gen  & 158 \\
			\hline
			20.   & Skintrim.N  & 80 \\
			\hline
			21.   &  Swizzot.gen!E  & 128 \\
			\hline
			22.   &  Swizzot.gen!I & 132 \\
			\hline
			23.   &  VB.AT  & 408 \\
			\hline
			24.   &  Wintrim.BX  & 97 \\
			\hline
			25.   &  Yuner.A & 800\\ 
			\hline
		\end{tabular}
	\end{table}

		\begin{table}[!htb]
	\captionsetup[table]{skip=10pt}
	\caption{Model Parameters for Classifier}\label{tbl:tab5}
	\centering    
	\begin{tabular}{|m{2cm}<{\centering}|m{3cm}<{\centering}|m{2cm}<{\centering}|}
		\hline
		Parameters & Implications  & Values \\ 
		\hline
		$w$    & width of image & 64\\ 
		\hline
		$h$     & height of image  & 64 \\ 
		\hline
		$f1$    & num of filters & 64\\ 
		\hline
		$conv1$     &   kernel size  & 5 \\ 
		\hline
		$l2$     &  kernel regulizer  & 0.01 \\ 
		\hline
		$st$   & stride  & 1 \\ 
		\hline
		$ps$  & pooling size  & 2 \\ 
		\hline
		$f2$   & num of filters  & 128 \\
		\hline
		$conv2$   & kernel size  & 5 \\
		\hline
		$f3$   & num of filters & 256 \\
		\hline
		$conv3$   & kernel size   & 2 \\
		\hline
		$r$   & drop rate  & 0.5 \\
		\hline
		$fc1$   & num of neurons  & 256 \\
		\hline
		$fc2$   & num of neurons & 128 \\
		\hline
		$fc3$   &  num of taxonomies  & 25 \\
		\hline
	\end{tabular}
\end{table}

	\textbf{Parameters.} In order to identify a proper structure of Classifier, we attempted the number of layers from 1 to 20. It turned out that a structure with $12$ layers is the most effective one considering the trade-off between accuracy and efficiency when Classifier without Transformation is employed to distinguish malware images. More specifically, the constructed classifier includes $4$ convolutional layers, 4 max-pooling layers, 1 dropout layer, and 3 fully connected feed-forward layers. The details of the model parameters are presented in Table \ref{tbl:tab5}.

	Once the structure of Classifier is fixed, we next determine the values of the following parameters defined in Section~\ref{sec:engineer}, which are used by the Transformation phase, in order to attain a better performance: $a$ and $b$, the numbers of pieces split up respectively for the width and height of a malware image during the Transformation phase; $clipLimit$, the upper bound of the frequency of a pixel value adopted by the Clipping phase; and $s$, the number of pixels in the width and height  of the transformed malware image to be inputted to Classifier at the end of Transformation. In our experiments, $clipLimit$ is empirically set to be 4 according to \cite{ahe}, and the shape $s$ of a transformed image is set to be 64 (the least image width) concerning the trade-off between accuracy and efficiency since a too small value of $s$ incurs a loss of image information while a too big value is bad for the fast processing of Classifier. 
	
	
  The determination of $a$ and $b$ is more complicated. Since different programs were developed for different purposes and thus have different sets of instructions to support their functionality, we consider the distribution of the instruction length to determine the value of $a$. In \cite{instrudis}, Ibrahim \emph{et al.} counted instruction length distributions for different types of programs, whose values were used to get the statistical information shown in Table \ref{tbl:tab3}. One can see that, for all types of programs, 64-bit executables and DLLs have the largest weighted average instruction length, i.e., 3.1 bytes; thus we round up the length to 4 bytes to accommodate an entire instruction in most cases. Correspondingly, we fix the width of an image region for all malware images, which should be a multiple of 4 intuitively. As a result, the value of $a$ varies for different malware images due to various image widths. Next we need to figure out the value of $b$. For simplicity, we choose a fixed $b$ determined by the average height of all malware samples belonging to the same family even though they may own different file sizes whose discrepancies are huge. Nevertheless,  a big value of $b$ (too many pieces in height) separates similar instruction sequences into different image regions, while a small value decreases the effectiveness of Transformation to enhance the local contrast at each image region.  Based on the above considerations, we tested the classification accuracy of all images in our dataset and attempted values of 8, 16, 32, 48, and 64 for the width of an image region, and values of 20 to 24 for the value of $b$, and selected the value of 32 and 23 respectively for the width of an image region and $b$. The detailed settings of $a$ and $b$ are listed in Table~\ref{tbl:tab4}.

		\begin{table}[!htb]
		\captionsetup[table]{skip=10pt}
		\caption{Weighted Average Instruction Length for Each Application Category}\label{tbl:tab3}
		\centering
		\begin{tabular}{|m{3.5cm}<{\centering}|m{3.5cm}<{\centering}|}
			\hline
			Application Category   &   Weighted Average Instruction Length  \\ 
			\hline
			Web Browsers           &   3.0                  \\ 
			\hline
			Graphics Applications  &   2.3              \\ 
			\hline
			OS related components  &   2.7                \\ 
			\hline
			General Purpose Applications & 2.9              \\ 
			\hline
			Software Development Tools &  2.5            \\
			\hline
			32-bit Executables and DLLs & 2.4            \\
			\hline
			64-bit Executables and DLLs& 3.1             \\
			\hline
			sum & 2.5 \\
			\hline
		\end{tabular}
	\end{table} 
	\begin{table}[!htb]
		\captionsetup[table]{skip=10pt}
		\caption{The Distribution of Malware Sample File Sizes}\label{tbl:tab4}
		\centering
		\begin{tabular}{|c|c|c|c|c|c|}
			\hline
			File Size  &   num\_of\_files &Width & a & Height & b\\ 
			\hline
			$\le$ 10KB   &   0  &  32 &0 &(0,320] &0  \\ 
			\hline
			10KB-30KB  &    701  & 64 &2 &(156,468] & 23        \\ 
			\hline
			30KB-60KB  &   1901 & 128 &4 &(234,468] &23  \\ 
			\hline
			60KB-100KB &  3209  & 256 &8 &(234, 390] &23     \\ 
			\hline
			100KB-200KB &  1116 & 384 &12 &(260,520] &23\\
			\hline
			200KB-500KB &  1071 & 512 &16 &(390, 976] &23 \\
			\hline
			500KB-1000KB& 1324  & 768 &24 &(651,1302] &23\\
			\hline
			1000KB $\leq$&17  & 1024 &32 &(976, $\infty$) &23\\
			\hline
		\end{tabular}
	\end{table}

	\subsection{Evaluation Results}
	\subsubsection{Accuracy}
	\label{sec:accuracy}
	
	  \begin{table}[!htb]
	 	\captionsetup[table]{skip=10pt}
	 	\caption{Classification Report}\label{tbl:tab6}
	 	\centering    
	 	\begin{tabular}{|m{1.7cm}<{\centering}|m{1.2cm}<{\centering}|m{1.2cm}<{\centering}|m{1.2cm}<{\centering}|m{1.2cm}<{\centering}|}
	 		\hline
	 		Malware Family & Precision  & Recall & F1 Score & accuracy \\
	 		\hline
	 		Adialer.C     &  99.2   &   100  &    99.6  &  100  \\
	 		\hline
	 		Agent.FYI     &  97.8   &   100  &    98.8   & 100  \\
	 		\hline
	 		Allaple.A     &  99.6   &   99.9   &   99.7  &  99.9     \\
	 		\hline
	 		Allaple.L     &  99.9   &   99.9   &   100   &  99.9  \\
	 		\hline
	 		Alueron.gen!J &  99.5   &   98.0  &    98.7  &  98.0     \\
	 		\hline
	 		Autorun.K     &  9.4   &   9.4  &    9.4  &  9.4    \\
	 		\hline
	 		C2LOP.gen!g   &  84.5   &   80.5 &     78.3  &  80.5    \\
	 		\hline
	 		C2LOP.P       & 76.1   &   76.8   &   75.2   &  76.7  \\
	 		\hline
	 		Dialplatform.B  &  99.5  &   100  &    99.7  & 100\\
	 		\hline
	 		Dontovo.A     &   99.4   &  100    &  99.7  &  100     \\
	 		\hline
	 		Fakerean    &   99.2    &  98.3    &  98.8  &  98.4    \\
	 		\hline
	 		Instantaccess   &    99.8   &   100  &    99.9 & 100       \\
	 		\hline
	 		Lolyda.AA1     &   97.9    &   96.7    &   97.1  & 96.7     \\
	 		\hline
	 		Lolyda.AA2     &   97.1  &    97.8   &    97.2   & 97.8    \\
	 		\hline
	 		Lolyda.AA3     &   99.2     &  97.6    &   98.3  &  97.6    \\
	 		\hline
	 		Lolyda.AT       & 98.2     &  98.2    &  98.2   &  98.1   \\
	 		\hline
	 		Malex.gen!J      &  97.3     &  97.8     &  97.2 &  97.8   \\
	 		\hline
	 		Obfuscator.AD     &  100     &  100    &   100   &  100   \\
	 		\hline
	 		Rbot!gen      &  94.4      & 94.5    &   94.0   &  94.3   \\
	 		\hline
	 		Skintrim.N     &   87.8      & 90.0    &   88.8 &  90.0       \\
	 		\hline
	 		Swizzor.gen!E    &    52.8    &   47.6    &   46.1 & 47.7      \\
	 		\hline
	 		Swizzor.gen!I     &   51.1     &  43.3    &   37.0 & 43.2      \\
	 		\hline
	 		VB.AT     &   99.1     &  97.9    &   98.5   &97.8    \\
	 		\hline
	 		Wintrim.BX    &   99.1     &  97.9    &   98.5 &  97.9    \\
	 		\hline
	 		Yuner.A    &    89.5   &    100    &   94.6   &  100   \\
	 		\hline
	 		weighted avg  &     95.3  &    96.0   &   95.2 & 96.0     \\
	 		\hline
	 	\end{tabular}
	 \end{table}  

    \begin{table*}[!htb]
    	\captionsetup[table]{skip=10pt}
    	\caption{Confusion Matrix}\label{tbl:tab7}
    	\centering    
    	\begin{tabular}{|m{0.28cm}<{\centering}|m{0.28cm}<{\centering}|m{0.35cm}<{\centering}|m{0.35cm}<{\centering}|m{0.28cm}<{\centering}|m{0.28cm}<{\centering}|m{0.28cm}<{\centering}|m{0.28cm}<{\centering}|m{0.28cm}<{\centering}|m{0.28cm}<{\centering}|m{0.28cm}<{\centering}|m{0.28cm}<{\centering}|m{0.28cm}<{\centering}|m{0.28cm}<{\centering}|m{0.28cm}<{\centering}|m{0.28cm}<{\centering}|m{0.28cm}<{\centering}|m{0.28cm}<{\centering}|m{0.28cm}<{\centering}|m{0.28cm}<{\centering}|m{0.28cm}<{\centering}|m{0.28cm}<{\centering}|m{0.28cm}<{\centering}|m{0.28cm}<{\centering}|m{0.28cm}<{\centering}|}
    		\hline
    		122	&0	&0&	0&	0&	0&	0&	0&	0&	0&	0	&0	&0	&0	&0&	0&	0&	0&	0&	0&	0&	0&	0&	0&	0\\
    		\hline
    		0&	116	&0	&0	&0&	0&	0&	0&	0&	0&	0	&0&	0&	0&	0&	0&	0&	0&	0&	0&	0&	
    		0&	0&	0&	0  \\
    		\hline
    		0&	0&	2947& 0&	0&	0&	0&	0&	0&	0&	2&	0&	0	&0	&0&	0&	0&	0&	0&	1&	0&	0&	0&	0&	0 \\
    		\hline
    		1&	0&	0&	1590&	0&	0&	0&	0&	0&	0&	0&	0&	0&	0&	0&	0&	0&	0&	0&	0&	0&	0&	0&	0&	0\\
    		\hline
    		0&	0&	0&	0&	194&	0&	0&	0&	0&	0&	1&	0&	0&	0&	0&	0&	3&	0&	0&	0&	0&	0&	0&	0&	0\\
    		\hline
    		0&	0&	0&	0&	0&	10&	0	&0	&0&	0&	0&	0&	0&	0&	0&	0&	0&	0&	0&	0&	0&	0&	0&	0&	96\\
    		\hline
    		0&	0&	0&	0&	0&	0&	161&	8&	0&	0&	0&	0&	0&	0&	0&	0&	0&	0&	5&	0&	11&	15&	0&	0&	0 \\
    		\hline
    		1&	0&	3&	0&	0&	0&	11&	112&	0&	0&	0&	0&	0	&0	&0&	0&	0&	0&	3&	1&	2&	12&	1&	0&	0\\
    		\hline
    		0&	0&	0&	0&	0&	0&	0&	0&	177	&0&	0&	0&	0&	0	&0	&0&	0&	0&	0&	0&	0&	0&	0&	0&	0\\
    		\hline
    		0&	0&	0&	0&	0&	0	&0	&0&	0&	162	&0	&0	&0&	0	&0	&0	&0&	0&	0&	0&	0&	0&	0&	0&	0\\
    		\hline
    		0&	0&	1&	0&	0&	0&	2&	2	&0&	0&	375&	0&	0&	0&	0&	0&	0&	0&	0&	0&	0&	1&	0&	0&	0\\
    		\hline
    		0&	0&	0&	0&	0&	0&	0&	0&	0&	0&	0&	431	&0&	0&	0&	0&	0&	0	&0	&0&	0&	0&	0&	0&	0\\
    		\hline
    		0&	0&	0&	0&	0&	0&	0	&0	&0&	0&	0&	0&	206	&5&	0&	0&	2&	0&	0&	0&	0&	0&	0&	0&	0\\
    		\hline
    		0&	0&	0&	0&	1&	0&	0&	0&	0&	0&	0&	0&	3&	180	&0	&0&	0&	0&	0&	0&	0&	0&	0&	0&	0\\
    		\hline
    		0&	1&	0&	0&	1&	0&	0&	0&	0&	0&	0&	0&	1&	0&	120&	1&	0&	0&	0&	0&	0&	0&	0&	0&	0\\
    		\hline
    		0&	0&	0&	0&	0&	0&	0&	0&	0&	0&	0&	0&	1&	0&	1&	156&	0&	0&	0&	0&	0&	0&	1&	0&	0\\
    		\hline
    		0&	0&	0&	1&	0&	0&	0&	0&	0&	0&	0	&0&	0&	0&	0&	0&	133&	0&	0&	1&	0&	0&	1&	0&	0\\
    		\hline
    		0&	0&	0&	0&	0&	0&	0&	0&	0&	0&	0&	0&	0&	0&	0&	0&	0&	142&	0&	0&	0&	0&	0&	0&	0\\
    		\hline
    		0&	0&	1&	0&	0&	0&	5&	1&	0&	0&	0&	0&	0&	0&	0&	0&	0&	0&	149&	0&	0&	1&	0&	1&	0\\
    		\hline
    		0&	0&	8&	0&	0&	0&	0&	0&	0&	0&	0&	0&	0&	0&	0&	0&	0&	0&	0&	72&	0&	0&	0&	0&	0\\
    		\hline
    		0&	0&	0&	0&	0&	0&	9&	5&	0&	0&	0&	0&	0&	0&	0&	0&	0&	0&	1&	0&	61&	52&	0&	0&	0\\
    		\hline
    		0&	1&	0&	0&	0&	0&	11&	27&	0&	0&	0&	0&	0&	0&	0&	0&	0&	0&	0&	0&	35&	57&	1&	0&	0\\
    		\hline
    		0&	0&	2&	0&	0&	0&	0&	0&	1&	1&	0&	1&	0&	1&	0&	2&	0&	0&	0&	0&	1&	0&	399&	0&	0\\
    		\hline
    		0&	0&	1&	0&	0&	0&	0&	0&	0&	0&	0&	0&	0&	0&	0&	0&	0&	0&	1&	0&	0&	0&	0&	95&	0\\
    		\hline
    		0&	0&	0&	0&	0&	0&	0&	0&	0&	0&	0&	0&	0&	0&	0&	0&	0&	0&	0&	0&	0&	0&	0&	0&	800\\
    		\hline
    	\end{tabular}
    \end{table*}
    
    We first evaluate the accuracy of VisMal over the Malimg dataset and then compare VisMal with its variant, i.e., the version of VisMal without the Transformation phase, and a few other methods using different techniques. We conducted 10-fold stratified cross-validation to examine the accuracy of VisMal. This approach ensures an approximately equal proportion of each family present at each fold. Among the 10 folds, 9 is used for the training procedure and the rest is reserved for the test procedure. We repeated this process 10 times, with each fold used exactly once as the test fold, to ensure that all malware samples are used for testing purposes to check VisMal's generalization ability. Finally, the average accuracy, precision, recall, and F1 score are computed for all malware families and the results are reported in Table \ref{tbl:tab6}.
   
\begin{figure}
    	\centering
    	\begin{subfigure}[b]{0.24\textwidth}
    		\centering
    		\includegraphics[height=1.4in,width=\textwidth]{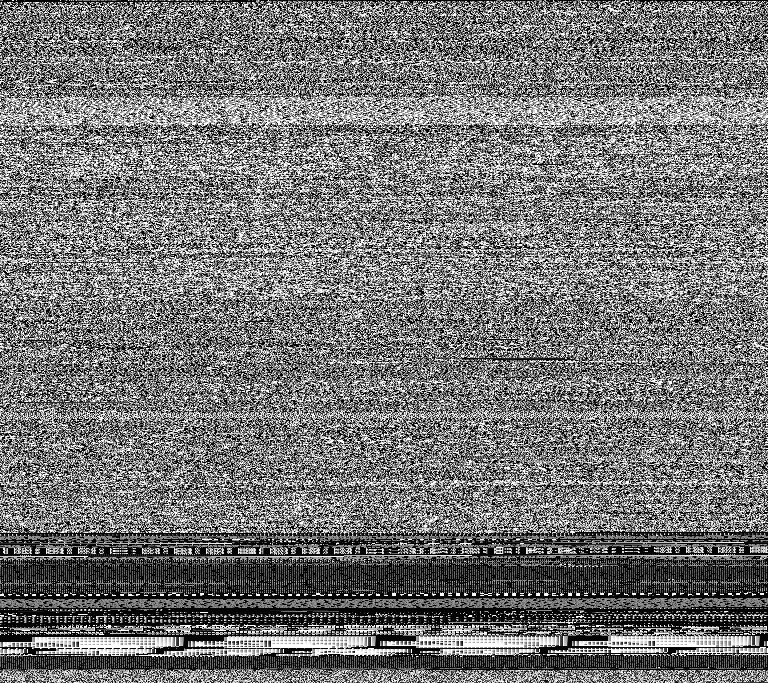}
    		\caption{Autorun.K sample1}
    		\label{fig:Autorun.K1}
    	\end{subfigure}
    	\begin{subfigure}[b]{0.24\textwidth}
    		\centering
    		\includegraphics[height=1.4in,width=\textwidth]{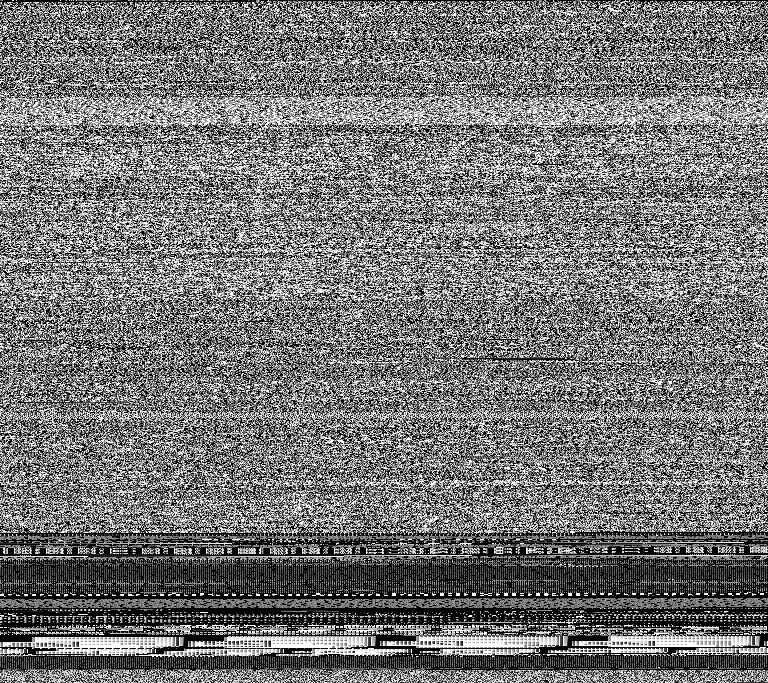}
    		\caption{Autorun.K sample2}
    		\label{fig:Autorun.K2}
    	\end{subfigure}
    	\hfill
    	\\
    	\begin{subfigure}[b]{0.24\textwidth}
    		\centering
    		\includegraphics[height=1.4in,width=\textwidth]{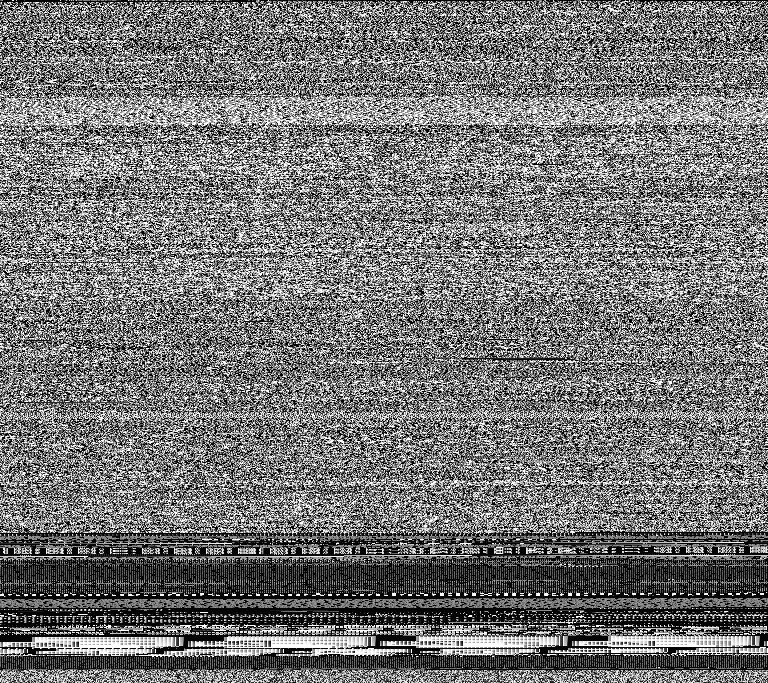}
    		\caption{Yuner.A sample1}
    		\label{fig:Yuner.A1}
    	\end{subfigure}
    	\begin{subfigure}[b]{0.24\textwidth}
    		\centering
    		\includegraphics[height=1.4in,width=\textwidth]{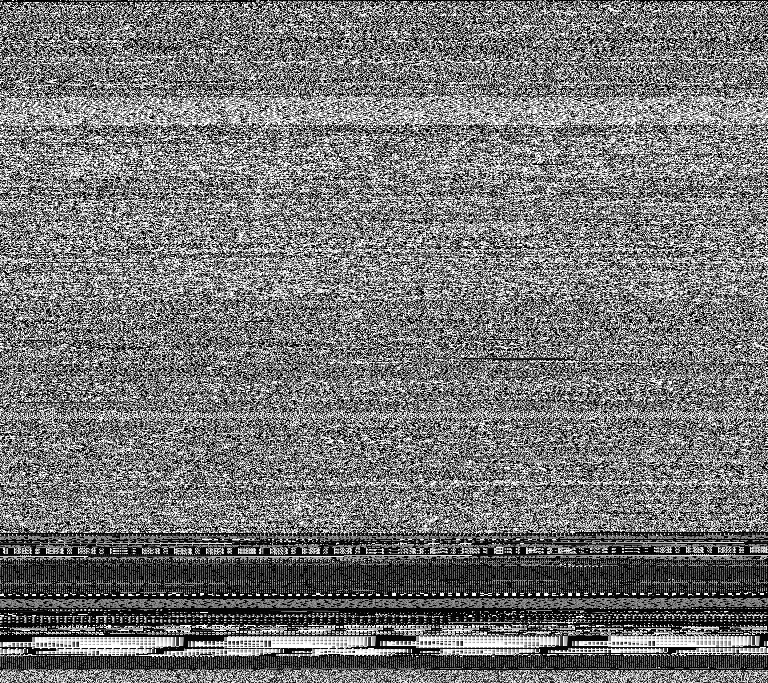}
    		\caption{Yuner.A sample2}
    		\label{fig:Yuner.A2}
    	\end{subfigure}
    	\\
    	\caption{Visual Similarity Example 1}
    	\label{fig:Yuner}	
    \end{figure}

\begin{figure}
    	\centering
    	\begin{subfigure}[b]{0.24\textwidth}
    		\centering
    		\includegraphics[height=1.4in,width=\textwidth]{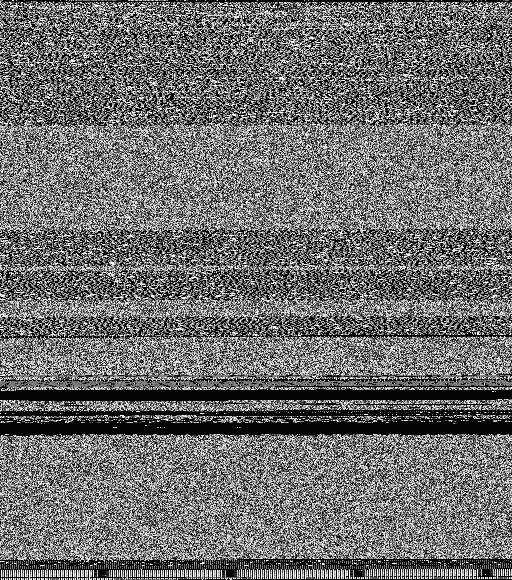}
    		\caption{Swizzor.gen!E sample1}
    		\label{fig:Swizzor.gen!E96e5}
    	\end{subfigure}
    	\begin{subfigure}[b]{0.24\textwidth}
    		\centering
    		\includegraphics[height=1.4in,width=\textwidth]{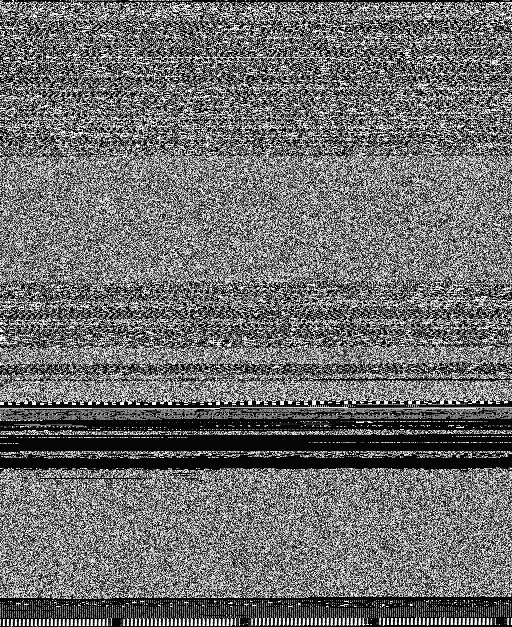}
    		\caption{Swizzor.gen!E sample2}
    		\label{fig:Swizzor.gen!Ecf12}
    	\end{subfigure}
    	\hfill
    	\\
    	\begin{subfigure}[b]{0.24\textwidth}
    		\centering
    		\includegraphics[height=1.4in,width=\textwidth]{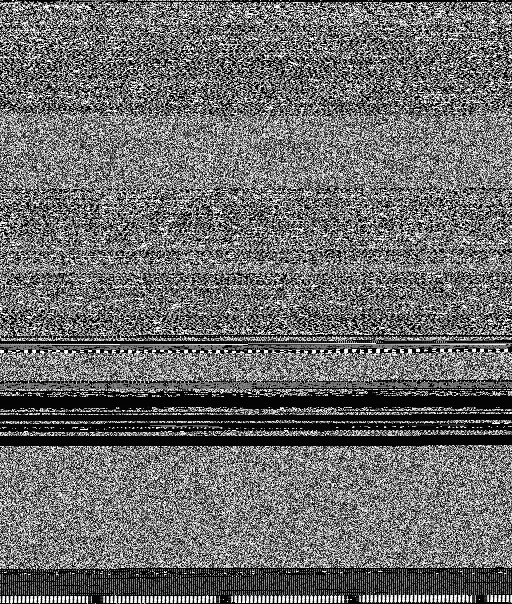}
    		\caption{Swizzorgen!I sample1}
    		\label{fig:Swizzor.gen!I5616}
    	\end{subfigure}
    	\begin{subfigure}[b]{0.24\textwidth}
    		\centering
    		\includegraphics[height=1.4in,width=\textwidth]{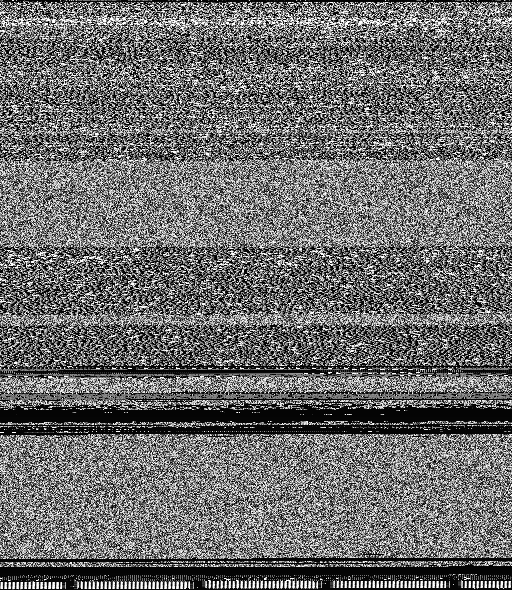}
    		\caption{Swizzor.gen!I sample2}
    		\label{fig:Swizzor.gen!Ia459}
    	\end{subfigure}
    	\\
    	\caption{Visual Similarity Example 2}
    	\label{fig:Swizzor}	
    \end{figure}

    One can see from Table \ref{tbl:tab6} that the average accuracy, precision, recall, and F1 score of VisMal are 96.0\%, 95.3\%, 96.0\% and 95.2\%, respectively. Specifically, among the 25 malware families, most of them are characterized correctly, with an accuracy close to 98\% or above. Note that some of the malware samples, including Yuner.A, VB.AT, Malex.gen!J, Autorun.K, and Rbot!gen in our dataset,  possess similar structures and patterns, and are packed by the UPX packer; thus it is difficult for VisMal to separate them.  But VisMal still achieves a promising result for all of them except Autorun.K. Specifically, VisMal classifies Yuner.A with a recall of 100\%, VB.AT of 97.9\%, Malex.gen!J of 97.8\%, and Rbot!gen of 94.5\%. Through a manual analysis, we found that the converted images of Autorun.K are extremely similar to those of Yuner.A. They are hard to be distinguished by human eyes as demonstrated by Fig. \ref{fig:Yuner}. As a result,  all misclassified Autorun.K malware samples are classified into Yuner.A, as illustrated by the confusion matrix in Table \ref{tbl:tab7} that shows the specific classification results. It should also be noted that the code sections of Allaple.A and Allaple.L are encrypted at several layers using random keys; but they are classified correctly with a  high recall, i.e., over 99\% for both families. What's more, the samples that are the variants of the same family, i.e., Swizzor.gen!E and Swizzor.gen!I, are highly similar (see examples in Fig. \ref{fig:Swizzor}), making Classifier fail to differentiate them since they share common code sequences. Therefore, the corresponding malware samples are mutually categorized into each other's families, ending at relatively low accuracy.
    
     \begin{figure*}[!htb]
    	\centering
    	\includegraphics[height=120mm,width=160mm]{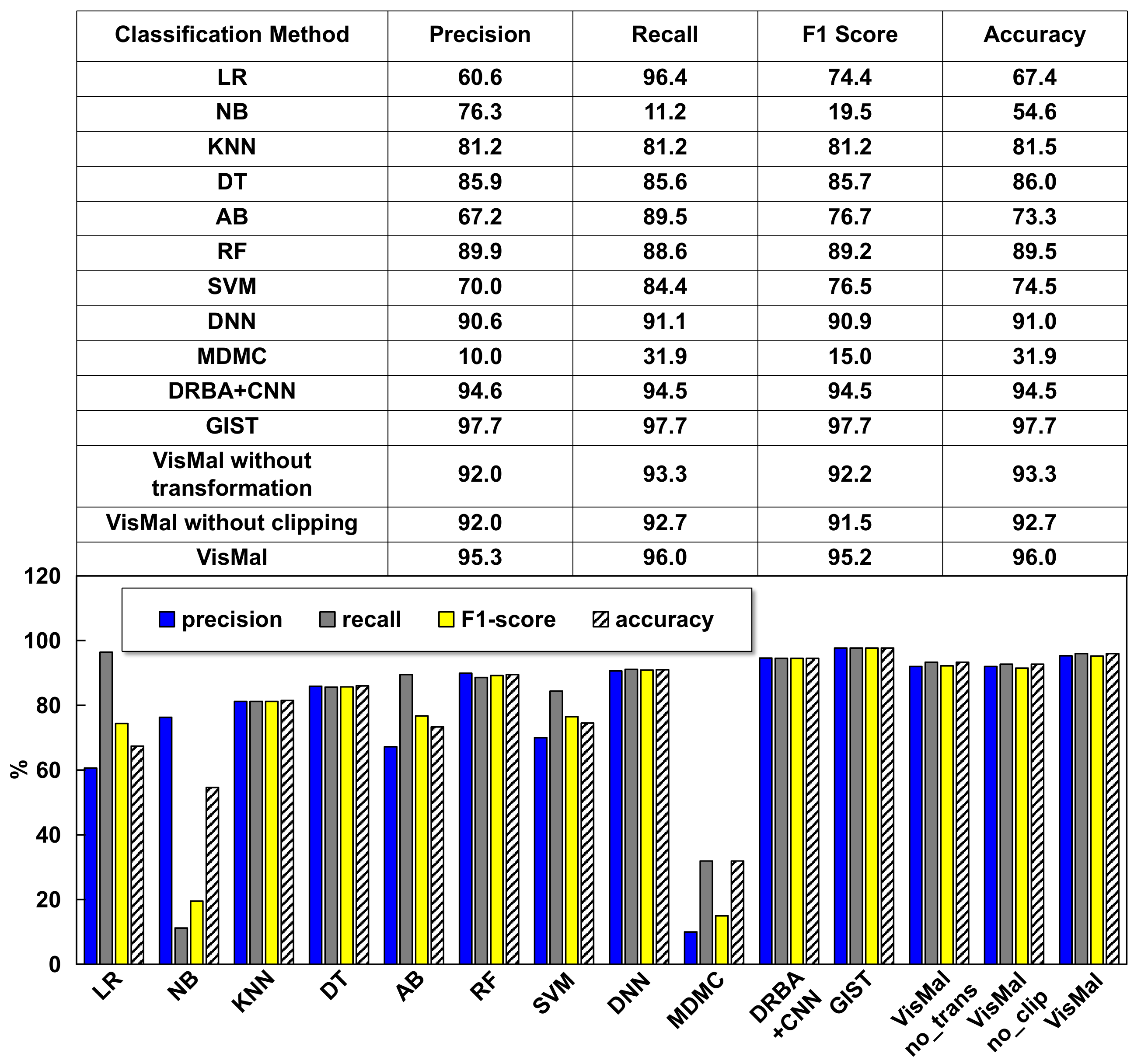}
    	\caption{Accuracy of the Classification Methods}
    	\label{fig:hist}
    \end{figure*}

    Fig. \ref{fig:hist} presents the comparison results among different classification techniques employed over the Malimg dataset \cite{robustin,cui,malimg,YUAN}. Compared to the VisMal variant without Transformation, VisMal improves its capability in recognizing different malware samples, which illustrates that the contrast for image regions is obliviously enhanced and is successfully captured by the developed convolutional neural network. Furthermore, the outstanding detection capability demonstrates that convolutional neural networks can utilize the property of many useful contents with compositional hierarchies in which higher-level features are obtained by composing lower-level ones. Compared to the VisMal variant without clipping, Vismal has a better performance on malware classification, which indicates that clipping is effective to offset the influence of noises due to the homogenous pixel distribution. Compared to other common machine learning-based and deep learning-based classification techniques, VisMal achieves a relatively higher accuracy of 96.0\%. To be specific, among the 7 common machine learning-based classification techniques and the deep learning-based approach presented in \cite{robustin},  the deep learning-based method demonstrates the best performance with an accuracy of 91\%, and LR achieves a very high recall at a moderate accuracy because the classifier categorizes a lot of malware samples from other families to the rest of them. The CNN-based framework proposed in \cite{cui}, namely DRBA+CNN, employs a BAT algorithm that can properly balance the number of samples in different malware families, achieving an accuracy of 94.5\%. Besides, we reimplemented two malware image-based frameworks, namely GIST and MDMC that were respectively proposed in \cite{malimg} and \cite{YUAN}, where  GIST utilizes the Gabor filters to extract statistical features for each malware gray image and carefully selects 320 features as the input to the $k$-nearest neighbor algorithm for final classification, while MDMC converts malware binaries into Markov probability matrixes by counting the frequency of two neighboring bytes and then employs a deep convolutional neural network to categorize different Markov probability matrixes. One can see that although GIST performs slightly better than VisMal in accuracy, its efficiency is far slower (described later). On other other hand, VisMal significantly outperforms MDMC in malware classification.

		\begin{table}[!htb]
		\captionsetup[table]{skip=10pt}
		\caption{Efficiency of the Classification Methods}\label{tbl:tab8}
		\centering    
		\begin{tabular}{|m{3cm}<{\centering}|m{1.2cm}<{\centering}|m{1.5cm}<{\centering}|m{1.2cm}<{\centering}|}
			\hline
			                                   & Extraction Time & Classification Time & Total Time \\ \hline
			Nataraj \emph{et al.}\cite{malimg} & 32.7 ms         & 2.1 ms              & 34.8 ms    \\ \hline
			Cui \emph{et al.} \cite{cui}       & --              & --                  & 20 ms      \\ \hline
			Naeem \emph{et al.} \cite{naeem}   & --              & 4.27 s              & --         \\ \hline
			Yuan \emph{et al.} \cite{YUAN}     & 144.3 ms        & 191.5 ms            & 335.8 ms   \\ \hline
			Vasan \emph{et al.} \cite{VASAN}   & --              & --                  & 1.18 s     \\ \hline
			Verma \emph{et al.} \cite{multi}   & 37 ms           & 10 ms               & 47 ms      \\ \hline
			VisMal                             & 0.3 ms          & 3.7 ms              & 4.0 ms     \\ \hline
		\end{tabular}
	\end{table}

	\subsubsection{Efficiency}
	\label{sec:efficiency}
	Since malware classification is a time-sensitive component in any anti-virus product, and a small delay could miss the best opportunity to discover malware processes, the procedure for distinguishing malware samples should take short time. We measured the CPU time of our dataset passing through VisMal during the test procedure and calculated the average MPE per malware sample. To consider a practical setting for evaluating VisMal, we did not deploy it on servers or call GPUs. The efficiency evaluation results are reported in Table \ref{tbl:tab8}. One can see that VisMal incurs 0.3 ms on feature extraction and 3.7 ms on classification, efficiently providing information to anti-virus products for further operations. Besides, we compared VisMal with the techniques presented by the latest research  (\cite{malimg} in 2011, \cite{cui} in 2018, \cite{naeem} in 2019, \cite{YUAN} in 2020, \cite{VASAN} in 2020, and \cite{multi} in 2020) that mainly pay attention to efficiency on malware classification over the same dataset. From Table \ref{tbl:tab8} we notice that VisMal takes the least extraction time and total time, and the second least classification time. Specifically, in terms of the time taken to extract features,  VisMal is 108 times faster than the method proposed by Nataraj \emph{et al.}, and 122.33 times faster than the method presented by Verma \emph{et al.} Additionally, the classification time incurred by VisMal is reduced by 1153.05 times and 1.70 times compared with the approaches in Naeem \emph{et al.} and Verna \emph{et al.}, respectively. Besides the extraction time and classification time, a significant decrease in total time (4 times) is reported for VisMal, when compared to Cui \emph{et al.}, the second-fastest method reported in Table \ref{tbl:tab8}.
	
   	\begin{figure}
   	\centering
   	\begin{subfigure}[b]{0.24\textwidth}
   		\centering
   		\includegraphics[height=1.4in,width=\textwidth]{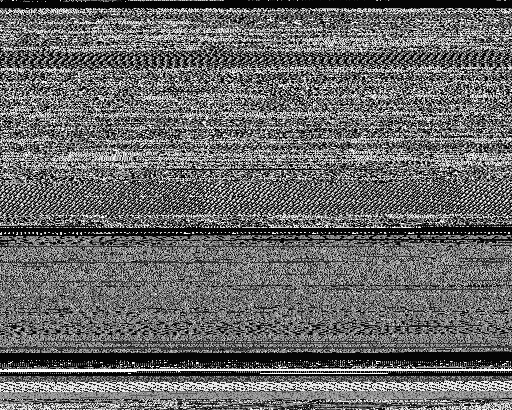}
   		\caption{Adialer.C sample1}
   		\label{fig:Adialer.C7c2}
   	\end{subfigure}
   	\begin{subfigure}[b]{0.24\textwidth}
   		\centering
   		\includegraphics[height=1.4in,width=\textwidth]{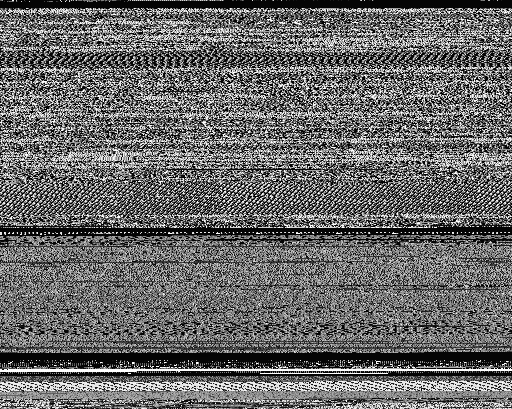}
   		\caption{Adialer.C sample2}
   		\label{fig:Adialer.C948}
   	\end{subfigure}
   	\hfill
   	\\
   	\begin{subfigure}[b]{0.24\textwidth}
   		\centering
   		\includegraphics[height=1.4in, width=\textwidth]{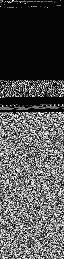}
   		\caption{Agent.FYI sample1}
   		\label{fig:Agent.FYI8ad}
   	\end{subfigure}
   	\begin{subfigure}[b]{0.24\textwidth}
   		\centering
   		\includegraphics[height=1.4in, width=\textwidth]{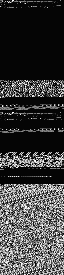}
   		\caption{Agent.FYI sample2}
   		\label{fig:Agent.FYIdf5}
   	\end{subfigure}
   	\\
   	\begin{subfigure}[b]{0.24\textwidth}
   		\centering
   		\includegraphics[height=1.4in, width=\textwidth]{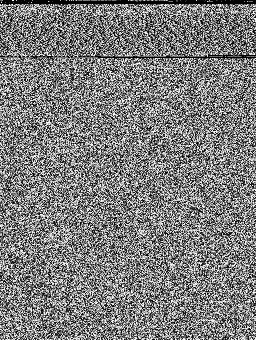}
   		\caption{Allaple.A sample1}
   		\label{fig:Allaple.A43ad}
   	\end{subfigure}
   	\begin{subfigure}[b]{0.24\textwidth}
   		\centering
   		\includegraphics[height=1.4in, width=\textwidth]{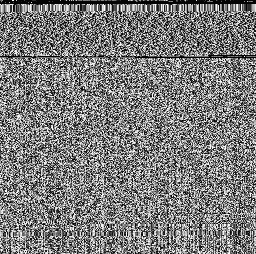}
   		\caption{Allaple.A sample2}
   		\label{fig:Allaple.Af4a}
   	\end{subfigure}
   	\\
   	\begin{subfigure}[b]{0.24\textwidth}
   		\centering
   		\includegraphics[height=1.4in, width=\textwidth]{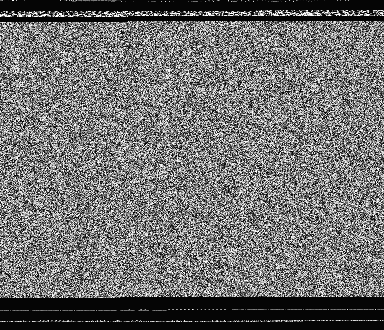}
   		\caption{Alueron.gen!J sample1}
   		\label{fig:Alueron.gen!Jbbb}
   	\end{subfigure}
   	\begin{subfigure}[b]{0.24\textwidth}
   		\centering
   		\includegraphics[height=1.4in, width=\textwidth]{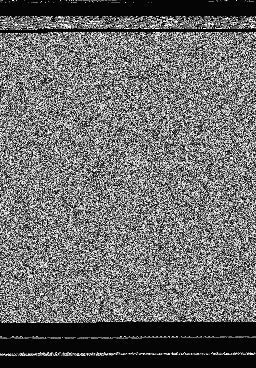}
   		\caption{Alueron.gen!J sample2}
   		\label{fig:Alueron.gen!Jd89}
   	\end{subfigure}
   	\\
   	\begin{subfigure}[b]{0.24\textwidth}
   		\centering
   		\includegraphics[height=1.4in, width=\textwidth]{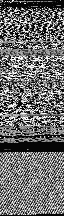}
   		\caption{Dialplatform.B sample1}
   		\label{fig:Dialplatform.B2df}
   	\end{subfigure}
   	\begin{subfigure}[b]{0.24\textwidth}
   		\centering
   		\includegraphics[height=1.4in, width=\textwidth]{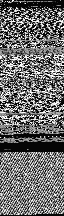}
   		\caption{Dialplatform.B sample2}
   		\label{fig:Dialplatform.Bce4}
   	\end{subfigure}
   	\\
   	\begin{subfigure}[b]{0.24\textwidth}
   		\centering
   		\includegraphics[height=1.4in, width=\textwidth]{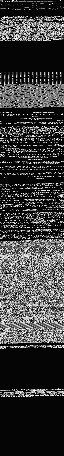}
   		\caption{Dontovo.A sample1}
   		\label{fig:Dontovo.A9da}
   	\end{subfigure}
   	\begin{subfigure}[b]{0.24\textwidth}
   		\centering
   		\includegraphics[height=1.4in, width=\textwidth]{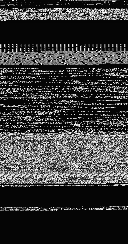}
   		\caption{Dontovo.A sample2}
   		\label{fig:Dontovo.A54a}
   	\end{subfigure}
   	\caption{Visual Similarity Example 3}
   	\label{fig:visualization}	
   \end{figure}
   
	\subsubsection{Visualization}
    \label{sec:visualization}
    Since malware classification is an important component integrated into all anti-virus tools, its precision plays an essential role in making these tools effective. To analyze the performance of VisMal, this section presents the visualization results in which malware samples are converted into images and then processed by the contrast-limited adaptive histogram equalization algorithm. Visualization provides useful information for manual analysis to ascertain whether the categorized malware samples are similar, which can help to confirm the performance of VisMal from a different angle. As VisMal converts malware samples into images, it naturally offers security engineers a convenient way to analyze malware samples and look for reasons when they are misclassified, based on which security engineers can integrate new techniques to improve the performance of their anti-virus products. 
    Fig. \ref{fig:visualization} demonstrates the texture similarity among malware samples belonging to the same family and dissimilarity from those belonging to different families. As shown in \ref{fig:Adialer.C7c2}-\ref{fig:Adialer.C948}, \ref{fig:Agent.FYI8ad}-\ref{fig:Agent.FYIdf5}, \ref{fig:Allaple.A43ad}-\ref{fig:Allaple.Af4a}, \ref{fig:Alueron.gen!Jbbb}-\ref{fig:Alueron.gen!Jd89},
    \ref{fig:Dialplatform.B2df}-\ref{fig:Dialplatform.Bce4},
    \ref{fig:Dontovo.A9da}-\ref{fig:Dontovo.A54a}, one can see that each pair is almost indistinguishable in terms of their texture information since, in the same family, the samples share certain common instruction sequences, and correspondingly their malware images tend to be similar in certain regions. Conversely, malware samples belonging to different families, which are developed for different purposes, make use of a dissimilar set of instruction sequences and thus have a huge difference in pixels' distribution, see Figs. \ref{fig:Adialer.C7c2} and \ref{fig:Alueron.gen!Jbbb} as an example.

	\section{Conclusion and Future Research}
	\label{sec:conclusion}
	In this paper, we propose a novel, efficient, and effective simple malware categorization framework titled VisMal, which can discover the underlying byte code similarity for malware samples that belong to a particular family. On one hand, VisMal alleviates the difficulty of security engineers to study the complicated structures of different malware samples based on static analysis, and in the meantime significantly improves the classifying efficiency from dynamic analysis. On the other hand, VisMal focuses on texture representation by converting malware samples into images, effectively subverting the traditional ways of understanding binary information.
	
	Note that VisMal pays attention to disclose the encoding art in byte codes and makes use of encoding to interpret the similarity of malware samples belonging to the same family and dissimilarity between different malware families. To improve the accuracy of classification, VisMal exploits a contrast-limited adaptive histogram equalization algorithm to increase the local contrast of each region for a malware image, and effectively transfers from looking for similar codes to similar image regions. Moreover, VisMal provides a simple and convenient way for security engineers to validate the performance of classification and to further analyze the factors confusing Classifier such that its performance can be improved by adding new techniques into anti-virus products. 
	
	Our future research will be committed to finding out the close relationship between the critical information of malware images learned by neural networks and the byte-code sequences, which may contribute to the extraction of signatures on behalf of a malware family. Such signatures can be managed via blockchain technologies \cite{wChain,BLOWN,CloudChain}  to facilitate malware detection by a broader audience.  We are also interested in the discovery of vulnerabilities in programs by making use of visualization techniques, which could provide new insights to help developers improve their codes for better classification performance. Besides, we will further strengthen our framework to distinguish malware and benignware.


\bibliographystyle{IEEEtran}
\bibliography{vismal}	
\end{document}